\pdfoutput=1
\documentclass[conference]{IEEEtran}
\IEEEoverridecommandlockouts 
\usepackage{cite}
\usepackage{amsmath,amssymb,amsfonts}
\usepackage{graphicx}
\usepackage{textcomp}
\usepackage{xcolor}
\usepackage{algorithm}
\usepackage{algpseudocode}
\def\BibTeX{{\rm B\kern-.05em{\sc i\kern-.025em b}\kern-.08em
		T\kern-.1667em\lower.7ex\hbox{E}\kern-.125emX}}
\usepackage{hyperref}

\begin{document}

\title{Katal\\
	\large A standard framework for finance
	\thanks{This work is licensed under the Creative Commons Attribution-ShareAlike 4.0 International License.}
}

\author{
	\IEEEauthorblockN{Bruno Fran\c{c}a, Sophie Radermacher and Reto Trinkler}
	\IEEEauthorblockA{
		Trinkler Software\\
		\textit{company@trinkler.software}}
	\\\\Version 3 -- July 3, 2019
}

\maketitle
\thispagestyle{plain} 
\pagestyle{plain} 

\begin{abstract}
Katal is a new blockchain that provides a standard way to build and deploy decentralized finance applications.

It brings together all the components necessary for the backend of a financial application, namely: a high-performance consensus, an authenticated data feed system, a standard for financial contracts and connectivity to the rest of the blockchain ecosystem.

Katal enables and simplifies the creation of financial services that are non-custodial, trustless, fast, convenient and interoperable.

\end{abstract}

\section{Introduction}
Katal aims to create a standard framework for the next stage of decentralized finance. In order to do this, it combines a high-performance blockchain, built using an innovative consensus algorithm and virtual machine, with a runtime that allows easy creation of assets, a standard for the creation of algorithmic financial contracts, and connectivity to the outside world and other blockchains.

Katal appeals to both developers and users:

\begin{itemize}
	\item For developers, Katal is a low-latency and high-throughput decentralized platform with native access to oracles, stable assets, connectivity to other blockchains and algorithmic contract templates that can be combined to create any type of financial contract. This allows developers to easily create decentralized financial services on top of Katal since custodianship, settlement and data feeds can be completely delegated to it.
	\item For users, Katal makes finance easy, using unprecedented speed, security, and risk management. Funds always remain under the user's control, empowering them to truly be their own banks. Settlement and data feeds are open and decentralized eliminating the need to trust exchanges. And, through a single Katal account, it is possible to interact seamlessly with all the financial services built on top of Katal (like futures and options exchanges, token exchanges, margin trading, short selling, collateralized loans, etc) and also with all the blockchains that connect to it.
\end{itemize}

In this paper we will introduce and describe every component of Katal and how they interact between themselves. The rest of the paper is divided as follows:

\begin{itemize}
	\item In section II we will discuss the main building blocks of Katal.
	\item In section III we will give an overview of the runtime and operation of Katal.
	\item In section IV we will delve into the runtime and describe it in detail.
	\item In section V we will explore several examples of applications that can be built on top of Katal.
\end{itemize}

\section{Building blocks}
Katal is built on top of an infrastructure that is composed of six pieces:

\begin{itemize}
	\item A blockchain development framework called Substrate,
	\item A consensus algorithm called Albatross,
	\item A virtual machine called Enso,
	\item A standard for the algorithmic description of financial contracts called ACTUS,
	\item An authenticated data feed system called Town Crier,
	\item And a heterogeneous multi-chain ecosystem called Polkadot.
\end{itemize}

We will now briefly explain each one of these components.

\subsection{Substrate}
\textit{Substrate} is a software framework for blockchains created by Parity Technologies. It packs a series of tools, written in Rust, that enables developers to easily create a blockchain.

For developers who prefer simplicity over freedom, it is possible to generate a new blockchain with just a simple configuration file. While developers that prefer freedom can create their own consensus algorithms and runtimes from scratch.

Substrate is a combination of mostly two technologies, WebAssembly and Libp2p. The built-in WebAssembly interpreter enables developers to write their own ready-to-use modules in any language they wish, as long as it compiles down to WebAssembly, and Libp2p constitutes the bulk of the networking functionality.

For Katal we will use the most bare-bones version of Substrate, called Substrate Core, that provides only the networking, the WebAssembly interpreter and some other auxiliary tools and we will provide the consensus and runtime modules.

\subsection{Albatross}
\textit{Albatross} \cite{franca2019albatross} is a novel consensus algorithm, developed by Trinkler Software and Nimiq, that is inspired by speculative BFT consensus algorithms.

Speculative BFT is a class of classical consensus algorithms that have very high performance when compared to older algorithms like PBFT \cite{castro1999practical}. Speculative BFT consensuses have two modes of operation: an \textit{optimistic} mode where nodes are assumed to be well-behaved thus resulting in greater performance, and the \textit{pessimistic} mode where it is assumed that some nodes are malicious and the only goal is to make progress.

Albatross takes a \textit{'trust but verify'} attitude to block production. Nodes are allowed to make updates to the state by themselves but other nodes can revert the update if it is not valid.

In Albatross there is a validator list that is selected at random from the set of all nodes that stake tokens. The validator list is changed every epoch, where an epoch is composed of \textit{T} micro blocks and one macro block.

\includegraphics[width=\linewidth]{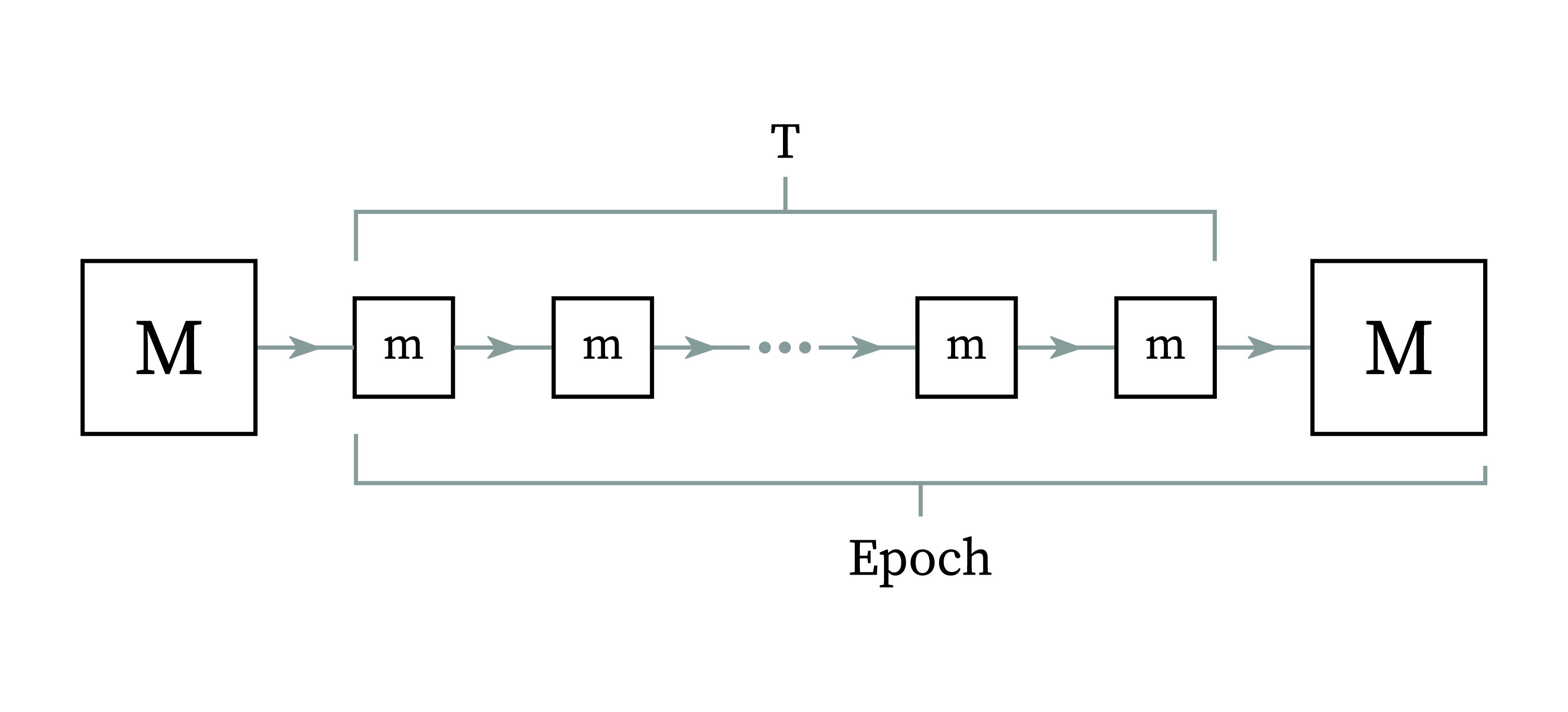}

Validators take turns producing micro blocks, which are transaction-containing blocks that are signed by a single validator. Macro blocks do not contain any transactions \footnote{We use same definition for transaction as the one taken by Parity. There are extrinsics, which are any input to the state transition function, and both transactions and inherents are mutually exclusive types of extrinsics. Transactions are extrinsics that are propagated through the network and signed. Inherents are neither propagated nor signed. An example of an inherent is a timestamp.}, being instead used to change the validator set, and are produced with PBFT.

If all validators follow the protocol correctly, Albatross will produce blocks as fast as the network allows. Since macro blocks don't contain any transactions, there will be a small pause at the end of every epoch, but that downtime constitutes only a small percentage of the total time. So, in the optimistic case, Albatross will come close to the theoretical limit for a single-chain consensus algorithm.

However, if there are malicious validators they may misbehave. There are three ways in which they can do that:

\begin{itemize}
	\item \textbf{Invalid blocks:} Validators may produce invalid blocks, in this case the other validators will simply ignore those blocks.
	\item \textbf{Forking:} Creating or continuing a fork will result in the stakes of the misbehaving validators being confiscated.
	\item \textbf{Delays:} If a validator, in his turn, does not produce a valid block after a predetermined amount of time, the other nodes will begin a \textit{view change} protocol that will allow another validator to produce a block.
\end{itemize}

Albatross has some desirable features besides its low latency and high throughput. It allows nodes to bootstrap quickly by only requiring them to download all of the macro blocks and the most recent micro blocks. It offers strong probabilistic finality, with a certainty of 99.9\% being reached in only 6 blocks. And honest clients can get instant confirmation that their transaction will be accepted by directly asking for receipts from the validators.

\subsection{Enso}
\textit{Enso} \cite{franca2019enso} is a general-purpose virtual machine for blockchains developed by Trinkler Software.

Blockchains can be seen as distributed virtual machines, since they are designed to run a given application on a network of heterogeneous, and possibly malicious, nodes. We can represent a blockchain as a stack:

\begin{center}
	\begin{tabular}{c c c}
		App data & $\equiv$ & State\\
		\cline{1-1} \cline{3-3}
		App logic & $\equiv$ & STF\\
		\cline{3-3}
		 &  & Consensus\\
		\cline{3-3}
		&  & Networking
	\end{tabular}
\end{center}

Where \textit{STF} is the state transition function. Note that the application logic, also called the \textit{runtime}, is embedded into the state transition function, while the application data is embedded in the state. For example, in the case of Bitcoin the application logic is that of a ledger and the application data is the set of all unspent transactions. This model works well for most cases but it is cumbersome to program and fails when it is necessary to update the application logic.

We take a different approach and have the state transition function be a general-purpose virtual machine. Then, the application logic and data can reside in the blockchain state:

\begin{center}
	\begin{tabular}{c c c}
		App logic \& data & $\equiv$ & State\\
		\cline{1-1} \cline{3-3}
		Virtual machine & $\equiv$ & STF\\
		\cline{3-3}
		 &  & Consensus\\
		\cline{3-3}
		&  & Networking
	\end{tabular}
\end{center}

Enso is this virtual machine. Having a general-purpose virtual machine speeds up development, since only the state needs to be programmed, and allows for simpler and more fine-grained updates, because the state can be changed with simple extrinsics instead of forks.

Enso itself is a relatively simple virtual machine, inside of it everything is either an \textit{object} or an \textit{event}. Objects are entities composed of:

\begin{itemize}
	\item \textbf{ID:} An unique identifier of the object. It can be any string.
	\item \textbf{Code:} A block of code containing functions that can be called by other objects.
	\item \textbf{Storage:} A data structure containing arbitrary information and that can be read and modified by the code.
\end{itemize}

While events are asynchronous function calls from one object to another and always have the following information:

\begin{itemize}
	\item \texttt{priority:} The priority of the event, used for the event queue.
	\item \texttt{ID\_to:} The ID of the object that will receive the event.
	\item \texttt{function\_call:} The name of the function that will be called.
	\item \texttt{parameters:} A list of parameters that will be passed to the function.
\end{itemize}

The state of the blockchain is the set of all objects, and it is these objects that will contain all the application logic and data. \textit{Everything is an object}.

The virtual machine itself is composed of just two components: the \textit{event queue} and the \textit{super object}.

The event queue is, as the name indicates, a queue for events. It is an ordinary priority queue and, when any event is created, it is added to this queue. Events with a higher priority go to the top of queue, while events with a lower priority go to the bottom.

The super object is a special object that is unique and can not be deleted or changed in any way. It is similar to the \textit{super user} in Linux systems, in that it has complete control over the state. In fact, it is the only object that can change the ID, code and storage of other objects. The super object has the following interface:

\begin{itemize}
	\item \texttt{Create(ID, code, storage):} Creates a new object with the given ID, code and storage.
	\item \texttt{Delete(ID):} Deletes object with the given ID.
	\item \texttt{Check(ID):} Checks if any object exists with the given ID and returns the answer.
	\item \texttt{Request\_object(ID):} Returns the code and storage of the object with the given ID.
	\item \texttt{Change\_ID(ID, new\_ID):} Changes the ID of the object with the given ID to \texttt{new\_ID}.
	\item \texttt{Change\_code(ID, new\_code):} Changes the code of the object with the given ID to new\_code.
	\item \texttt{Change\_storage(ID, new\_storage):} Changes the storage of the object with the given ID to new\_storage.
	\item \texttt{Set\_input(ID):} Sets the input object to the object with given ID.
\end{itemize}

The \textit{input object} is the object that is designated to receive extrinsics. An extrinsic in Enso is just a event like any other, but with the caveat that it must be sent to the input object. Any object can be the input object, the only requirement is that it designated as such by the super object.

A state transition in Enso looks like this:

\begin{enumerate}
	\item Receive an ordered list of extrinsics.
	\item Add the first extrinsic to the event queue.
	\item The resulting event will be sent to the input object. The input object then may create more events, that in turn will also be processed and create more events, and so forth, until no more events are created.
	\item Add the next extrinsic to the event queue and keep repeating steps 3 and 4 until there are no more extrinsics.
\end{enumerate}

\subsection{ACTUS}
The \textit{Algorithmic Contract Types Unified Standards} \cite{bundi2018actus}, or \textit{ACTUS}, is a standard developed by the ACTUS Financial Research Foundation. It seeks to describe all possible financial contracts as algorithmic patterns of cash flows between two parties.

In ACTUS, any financial contract can be replicated as a combination of simpler contracts, called \textit{contract types}. These contract types include basic financial contracts, like annuities and futures, and more exotic ones, like perpetual bonds and credit default swaps. In total there are 32 contract types, and together they form a taxonomy of financial contracts.

\begin{figure*}[t]
	\centering
	\includegraphics[width=\textwidth]{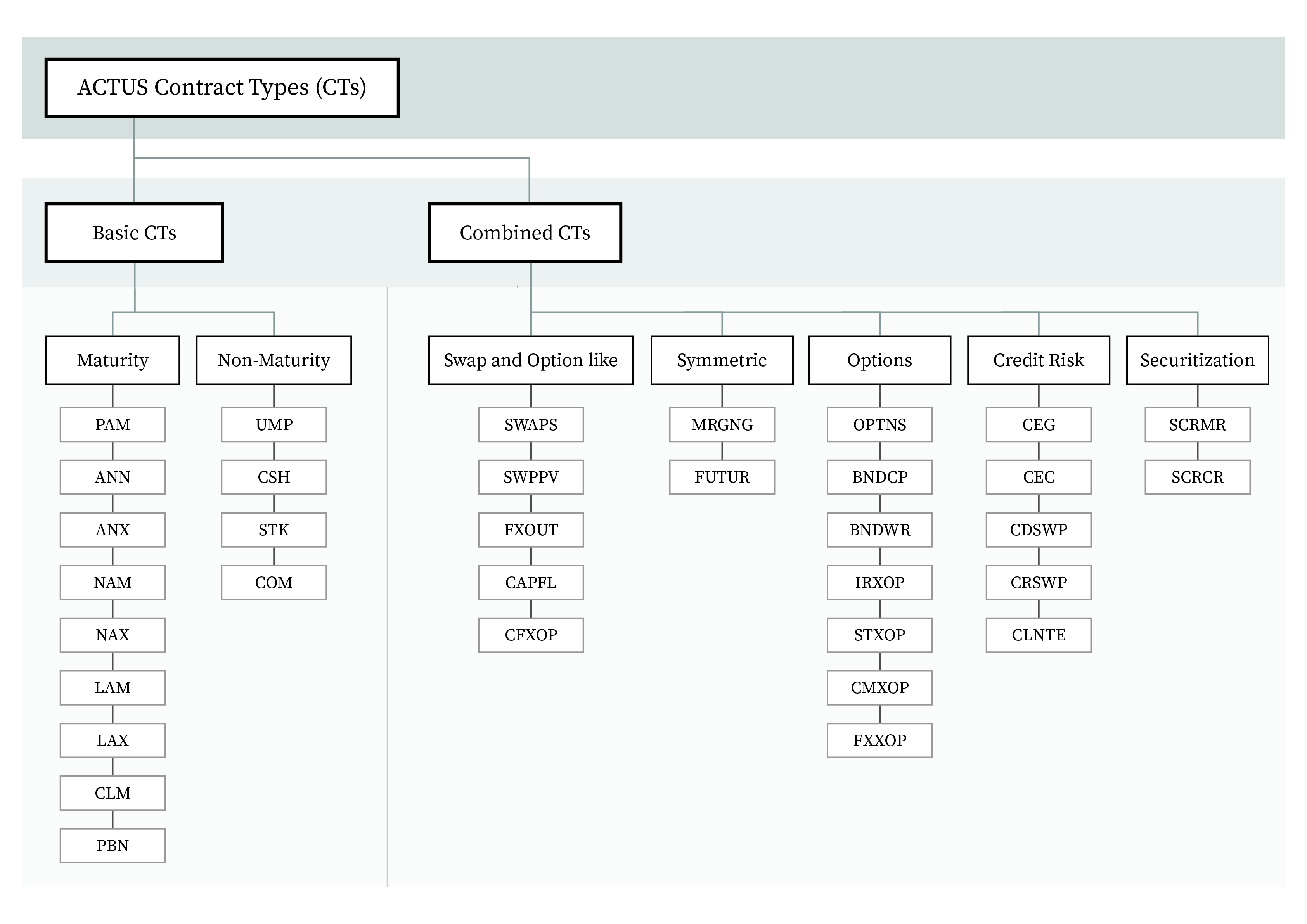}
	\caption{The ACTUS taxonomy of financial contracts.}
\end{figure*}

The contract types themselves are defined in algorithmic form. In other words, for each contract type there is a set of rules that, given some input parameters and external data, determine unambiguously the cash flows between the two parties to the contract. This allows them to be easily implemented in smart contracts. And, by combining different contract types, we can create any imaginable financial contract in algorithmic form.

There are five types of parameters that define a contract type: attributes, variables, contract events, payoff function and state transition function:

\begin{itemize}
	\item Attributes are parameters that are static. They are defined when the contract is created and then they never change.
	\item Variables, as the name indicates, are parameters that are dynamic. At the time of the contract creation they are initialized to a given value, but they may change afterwards.
	\item Contract events are actions that cause a cash flow and/or a change in the variables. They may be scheduled, for example at the beginning of every month, or initiated by an external object, for example by one of the parties.
	\item The payoff function takes the attributes, the current variables and an event as inputs, and outputs a cash flow obligation from one party to another. The state transition function takes exactly the same inputs, but it outputs new values for the variables. Together, these two functions constitute the logic of the contract.
\end{itemize}

Contracts have two interfaces that enable it to interact with other objects: the risk factor observer and the child contract observer.

The risk factor observer allows the contract to request data from an oracle object, such data can be, for example, price information or interest rates.

The child contract observer allows one contract to observe the attributes, variables or events of another contract. This functionality is what makes it possible for several contract types to be joined together into more complex financial contracts.

Another useful characteristic of ACTUS contracts is that the cash flows between both parties of a contract, called the \textit{creator} and the \textit{counterparty}, can be \textit{tokenized}. By this we mean that it is possible for users to have fractional ownership of a contract, thus it is perfectly possible to have the creator's or counterparty's cash flows (both positive and negative) be divided among several different users.

Tokenization also creates a simple way of transferring ownership of a contract. This feature is optional for any contract, since in some cases it may be undesirable, but when activated, it makes it possible for contracts to be sold in secondary markets.

\subsection{Town Crier}
\textit{Town Crier} \cite{zhang2016town}, developed by the Initiative for cryptocurrency and Contract (IC3), is an authenticated data feed system for smart contracts, also known as an \textit{oracle}. It acts as a bridge between HTTPS-enabled websites and blockchains, and does so by taking advantage of a \textit{trusted execution environment}, specifically the \textit{Intel SGX}.

A trusted execution environment can be thought of as a space inside the CPU that allows programs that run inside it to be protected from other programs, the operating system and even from some types of hardware attacks. This space is called an \textit{enclave}, and it is basically a \textit{black box} inside which programs are certain to be executed correctly and with confidentiality.

An enclave can only use the CPU and the RAM by itself and needs to rely on the operating system for file and network access. However, by using public key cryptography, it can establish secure connections over the internet.

Another useful feature of enclaves is that they can provide a publicly verifiable proof that a given program was executed correctly and produced a given output. Such a proof is called an \textit{attestation}.

In order to serve source-authenticated data to smart contracts, the Town Crier system only needs a specific smart contract, called the \textit{oracle}, and a relaying server, called the \textit{TC server}.

\begin{figure}[h]
	\includegraphics[width=\linewidth]{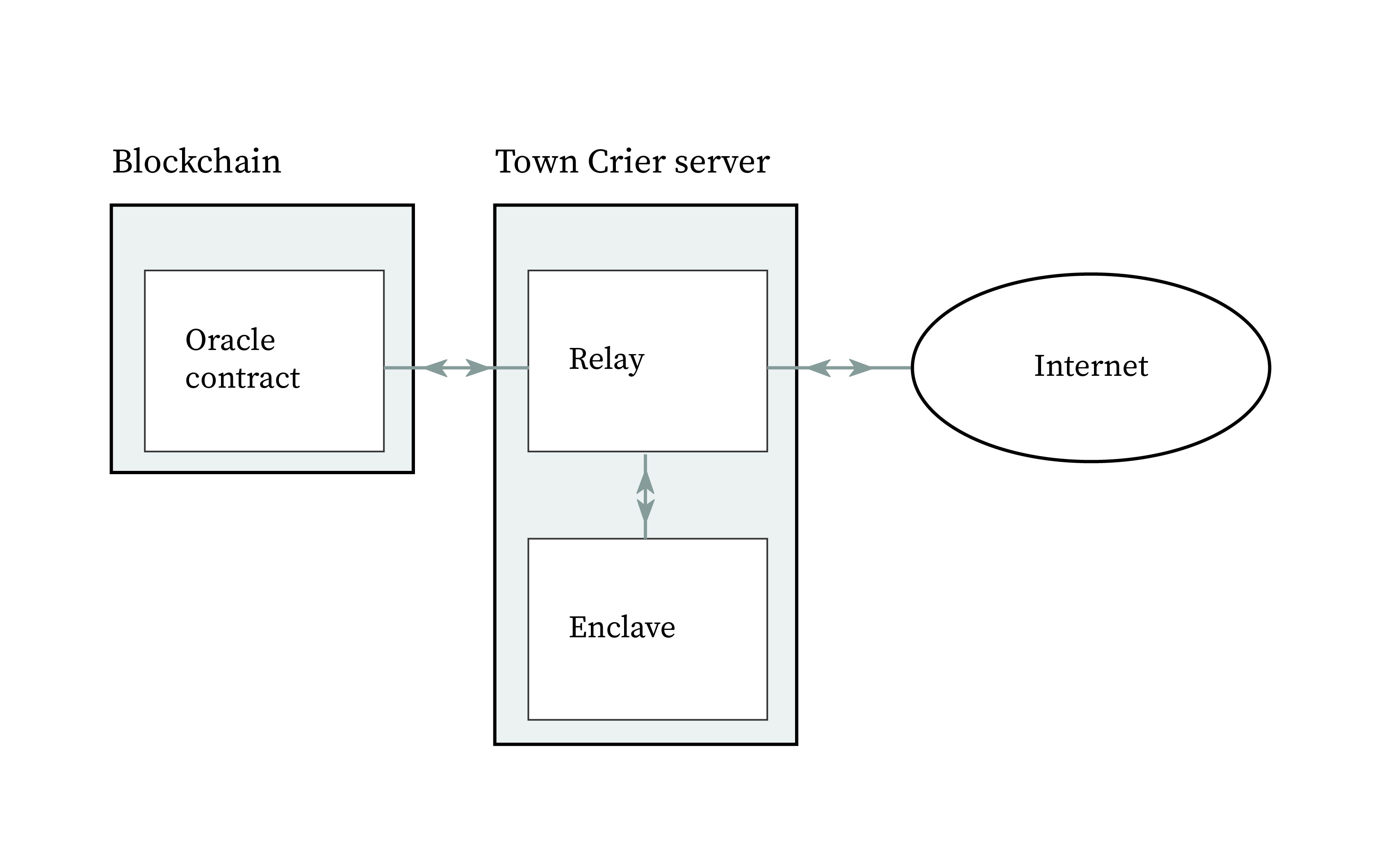}
	\caption{The Town Crier system.}
\end{figure}

The oracle contract acts as the front-end for the blockchain, creating requests for data, verifying attestations from enclaves and distributing rewards to servers that provide data.

The TC server has two components: the relay and the enclave. The relay handles all the network traffic to and from the enclave, since the enclave itself has no networking capabilities. The enclave establishes HTTPS connections to websites and produces attestations.

The process works as follows:

\begin{enumerate}
	\item The oracle contract produces a request for data. It does this by updating its state and signaling that it is ready to receive data from a TC server.
	\item The relay, who periodically watches the blockchain, sees the data request and relays it to the enclave.
	\item The enclave processes the request and initiates a HTTPS connection to the requested website.
	\item The relay handles the traffic between the enclave and the website during the HTTPS session.
	\item The enclave scrapes the website for the requested data and produces an attestation that the scraping was done correctly. Then, it sends the data and the attestation to the oracle.
	\item The relay forwards the data and attestation to the oracle.
	\item The oracle, after verifying that the attestation is correct, updates its state with the new data. Then, if appropriate, it distributes a reward to the TC server.
	\item At any point, other contracts can fetch data from the oracle by requesting its more recent state.
\end{enumerate}

Town Crier was recently acquired by Chainlink \cite{ellis2017chainlink}, a project that provides decentralized oracles for a variety of blockchains. Oracle contracts in Katal are updated by Chainlink nodes using the Town Crier protocol.

\subsection{Polkadot}
\textit{Polkadot} \cite{wood2016polkadot} is a heterogeneous multichain framework introduced by Gavin Wood in 2016. It is a protocol that allows blockchains to exchange information, but unlike internet messaging protocols like TCP/IP, Polkadot also enforces the order and the validity of the messages between the blockchains.

Polkadot is composed of a central blockchain, called the \textit{relaychain}, and a number of peripheral blockchains, called \textit{parachains}, that connect to it. Parachains may outsource their consensus to the relaychain or have their own consensus algorithm and validators. The relaychain, as the name indicates, acts as a relay for messages between different parachains.

\begin{figure}[h]
	\includegraphics[width=\linewidth]{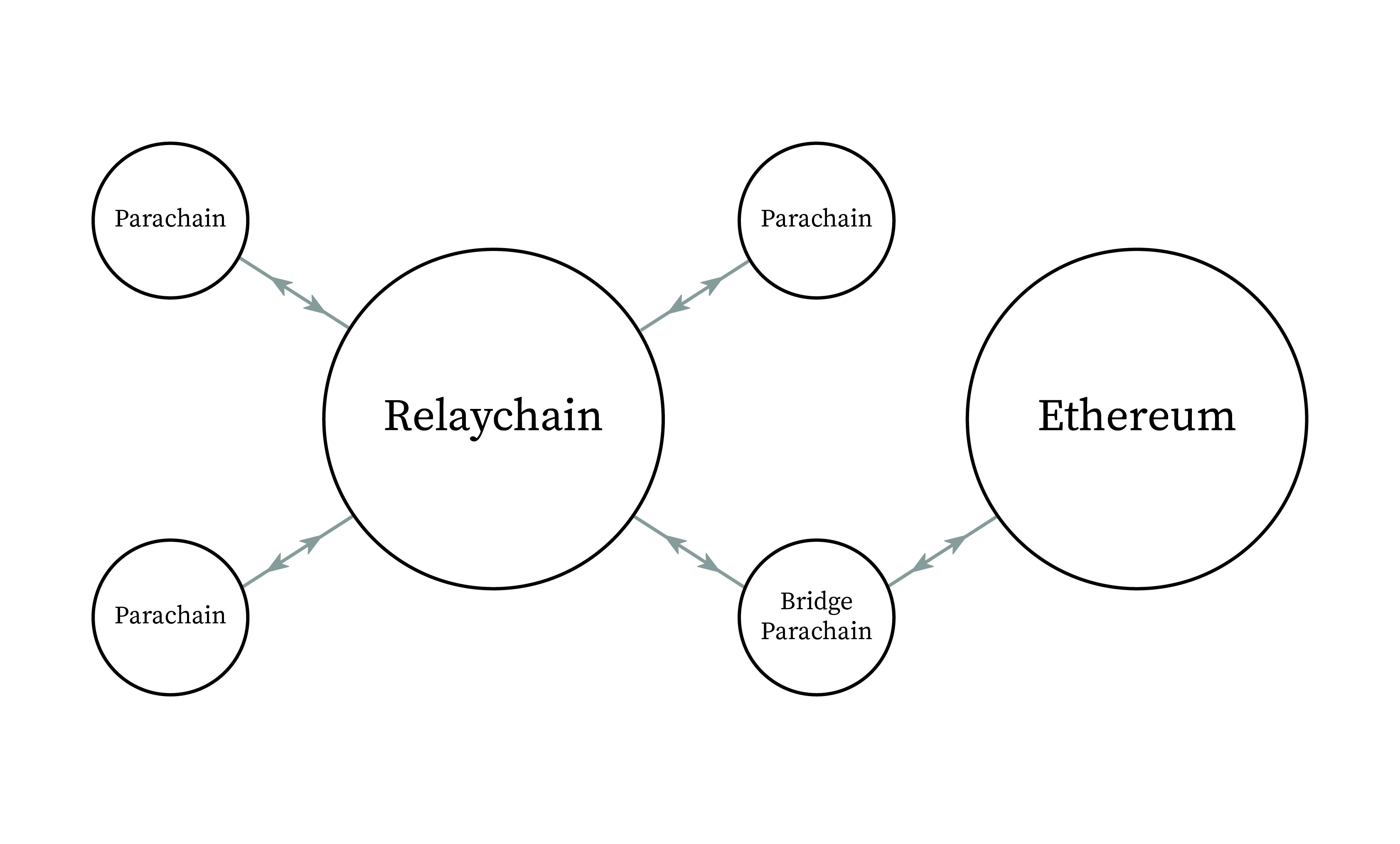}
	\caption{Polkadot: relaychain and parachains}
\end{figure}

Connecting to Polkadot enables Katal to not only exchange information with other parachains but also for other parachains to transfer tokens to the Katal blockchain and vice-versa.

\section{Overview}
\begin{figure}[h]
	\includegraphics[width=\linewidth]{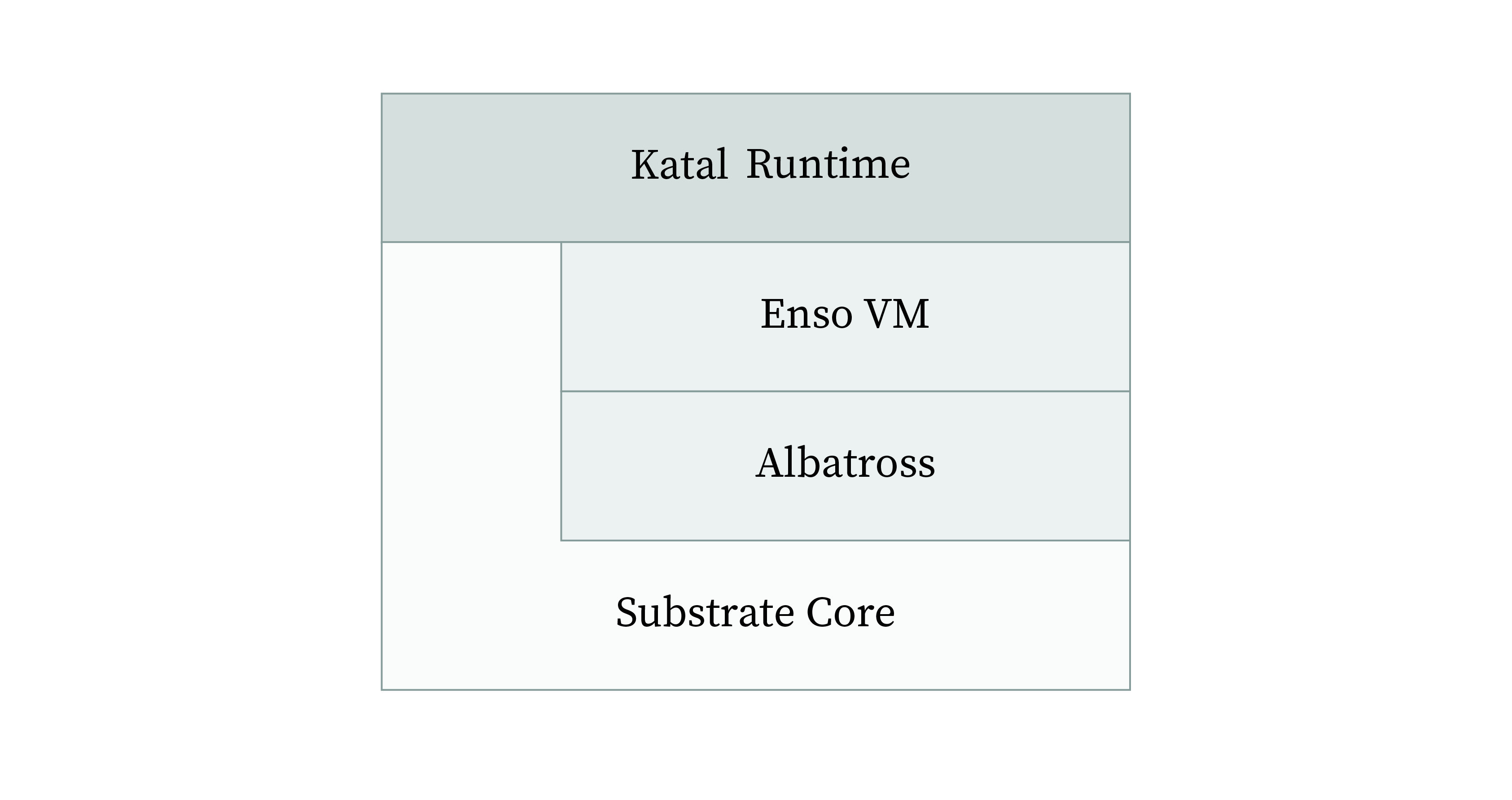}
	\caption{Katal technology stack}
	\label{fig:stack}
\end{figure}

The Katal technology stack is illustrated in Figure \ref{fig:stack}.

Analyzing it we see that Substrate Core is at the bottom and is used for networking, module swaps and other auxiliary services. Albatross and Enso are, respectively, the consensus algorithm and the virtual machine, and they will be implemented as modules in Substrate. Lastly, there is the Katal runtime on top.

It is the runtime that enables Katal capabilities and so, for the rest of this paper, we will focus mostly on it. Both Albatross and Enso are explained in detail in two other papers \cite{franca2019albatross} \cite{franca2019enso}.

The Katal runtime is composed of a set of objects and the interactions among them. Broadly speaking, the objects can be divided in two categories:

\begin{itemize}
	\item \textbf{Kernel objects:} All the objects that are created at the genesis block and during updates to the blockchain. These are the objects that define the rules for how the runtime works and are unique objects, meaning that there is only one instance of each object type. An example of a kernel object is the \textit{governance object}, which handles updates to the blockchain.
	\item \textbf{User objects:} All the objects that are created by the users. Any user can create user objects from a predetermined list of object templates. For example, there is an \textit{account template} and each instance of that template, created by the users, is an \textit{account object}.
\end{itemize}

For a full description of the Katal runtime it is enough to outline all the kernel objects plus all the different object templates. Let us begin with the kernel objects:

\begin{itemize}
	\item \textbf{Dock object:} It serves as the point of entry for extrinsics and is always the first object to be called, in other words it is the \textit{input object}. It parses each extrinsic, verifies their signatures and then creates the events to the desired objects.
	\item \textbf{Authentication object:} It maintains a list of the IDs of all user objects and their corresponding authentication methods. The dock object calls this object to verify the signatures on transactions. Other objects may also call it when they need authentication services.
	\item \textbf{Schedule object:} It enables periodic, or scheduled, calls to other objects. Every block it receives from the dock object the current time and block number and then proceeds by calling any objects that are scheduled to be called at that particular time.
	\item \textbf{Instantiation object:} It creates all user objects. Users can call it to create a new user object from a list of object templates. It maintains the list of object templates and defines which parameters are allowed for the instantiation of those templates.
	\item \textbf{Native issuance object:} It is a special instance of the more general \textit{issuance template}. It manages the native token (\textit{XTL}) by maintaining a list of all object IDs and their corresponding balance. It also deals with transferring, minting and burning \textit{XTL} tokens.
	\item \textbf{Governance object:} It implements whichever governance method is chosen to update the blockchain. Notably, it is the only object that possesses unrestricted access to the \textit{super object}, thus allowing it to modify any part of the Katal runtime.
	\item \textbf{Consensus object:} It manages certain tasks related to the consensus algorithm. Namely, it maintains a list of validators and their staking deposits, applies slash inherents, distributes block rewards, etc.
\end{itemize}

Besides these seven kernel objects, the Katal runtime is also composed of the following four user object templates:

\begin{itemize}
	\item \textbf{Account object:} It is the most basic object in the runtime, being basically just an ID. Whenever an account is created, a corresponding entry is created in the authentication object with the authentication method and parameters chosen by the user.
	\item \textbf{Issuance object:} It creates and manages its own tokens. It can be user-controlled or automated, is capable of managing several different token types and supports both fungible and non-fungible tokens.
	\item \textbf{Oracle object:} It maintains an external data feed by serving as the interface for Town Crier servers. It creates requests for data, validates the data authenticity and distributes rewards to servers.
	\item \textbf{Contract object:} There are actually 32 different contract templates, one for each ACTUS contract type. All contract objects are capable of interacting with issuance objects, to create cash flows, and with oracle objects, to fetch data. It also maintains a list of all its owners, both creators and counterparties, for tokenization purposes.
\end{itemize}

The above list gives a rough description of all the objects but it does not shed much light on how they interact together. Even though it is not possible to detail all possible interactions in this paper, we will now discuss some of the most common ones.

\subsection{Transaction docking}
All transactions pass first through the dock object. Any transaction is just a \textit{`bundle'} of function calls to other objects, and the dock object must first verify the validity of the transaction before creating the corresponding events for those function calls.

\includegraphics[width=\linewidth]{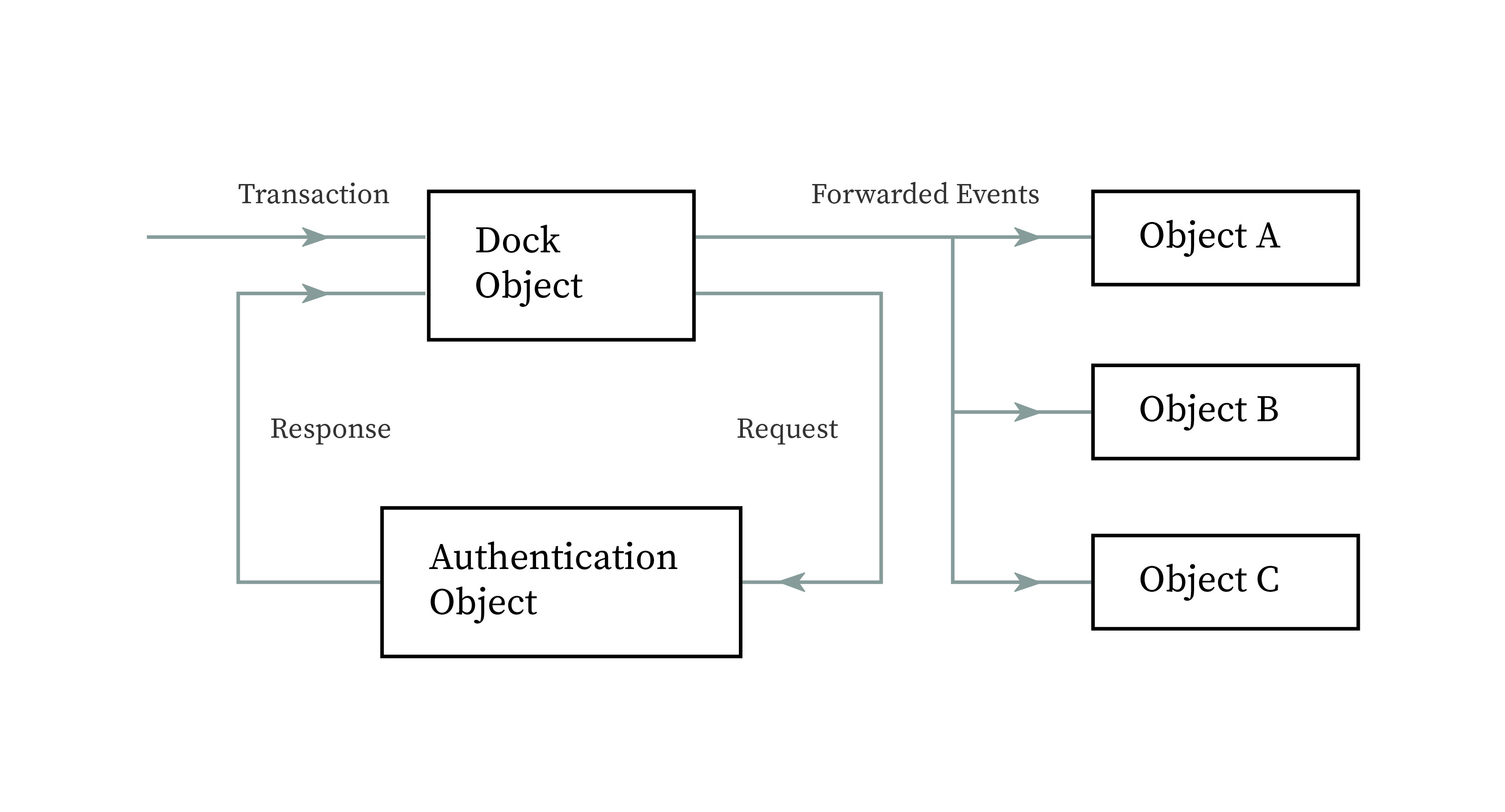}

When the dock object receives a transaction it will first check that it is well-formed. Then it will call the authentication object to verify the signatures contained in the transaction. If, and only if, all signatures are valid then the dock object will create the requested events. Each of these events will include, as part of their parameters, the IDs of everyone that signed the transaction.

\subsection{Token transfers}
Like ERC-20 tokens in Ethereum \cite{buterin2014next}, tokens in Katal are managed by a single object. In other words, instead of every account storing the balance of every token that it owns, for each token there is a single object that stores the ID and balance of every account that owns those tokens. These objects we call \textit{issuance objects}.

\includegraphics[width=\linewidth]{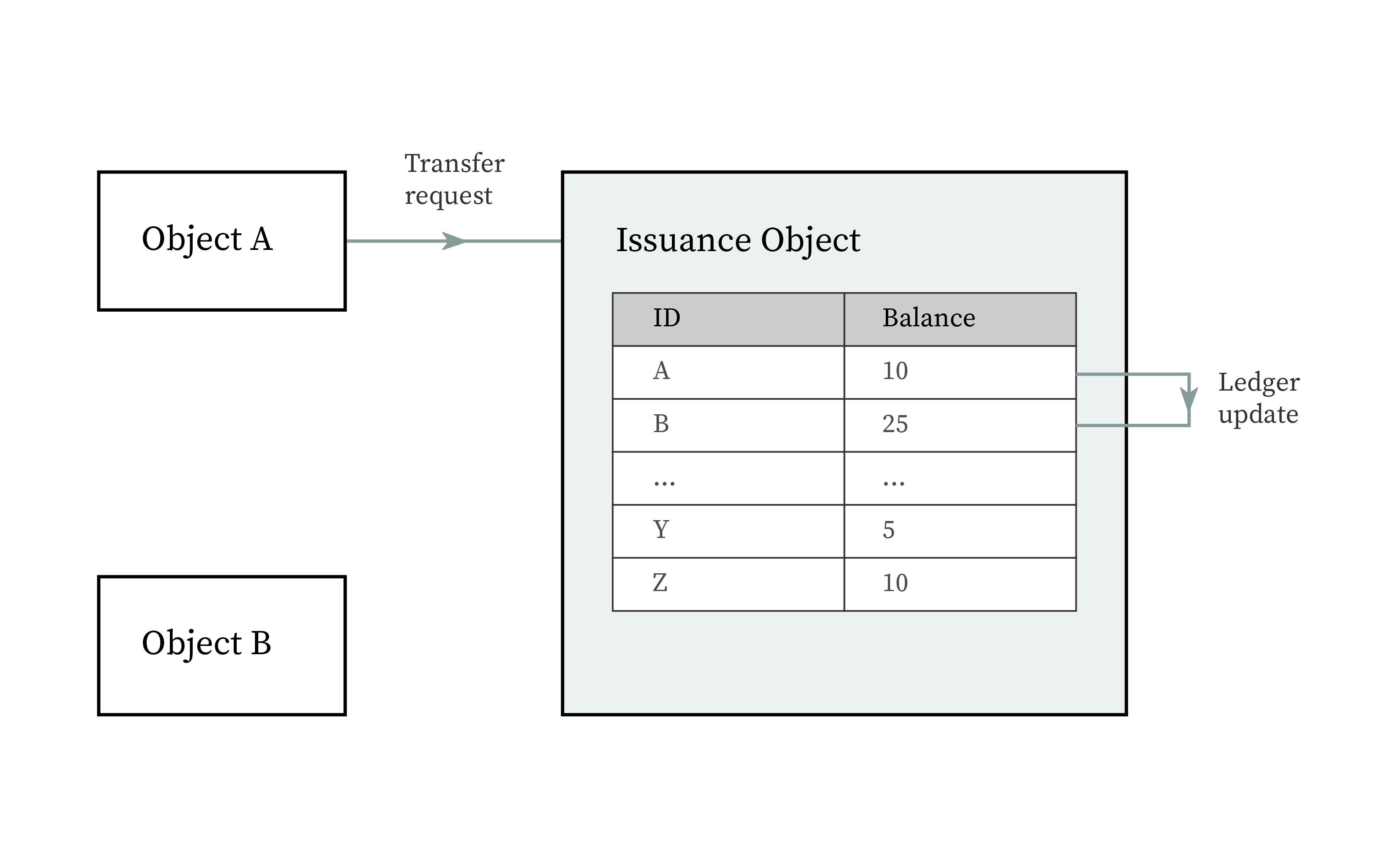}

When an account wants to transfer some tokens to another account, it just sends a request to the issuance object of that token. The issuance object will then update its internal ledger to reflect this change.

\subsection{Schedule functions}
It is useful to have functions that are started at predefined times. For example, we may want an oracle that is updated every minute. In order for this to happen, some object needs to call the oracle object every minute, so that it can accept updates to its state. That object is the schedule object.

\includegraphics[width=\linewidth]{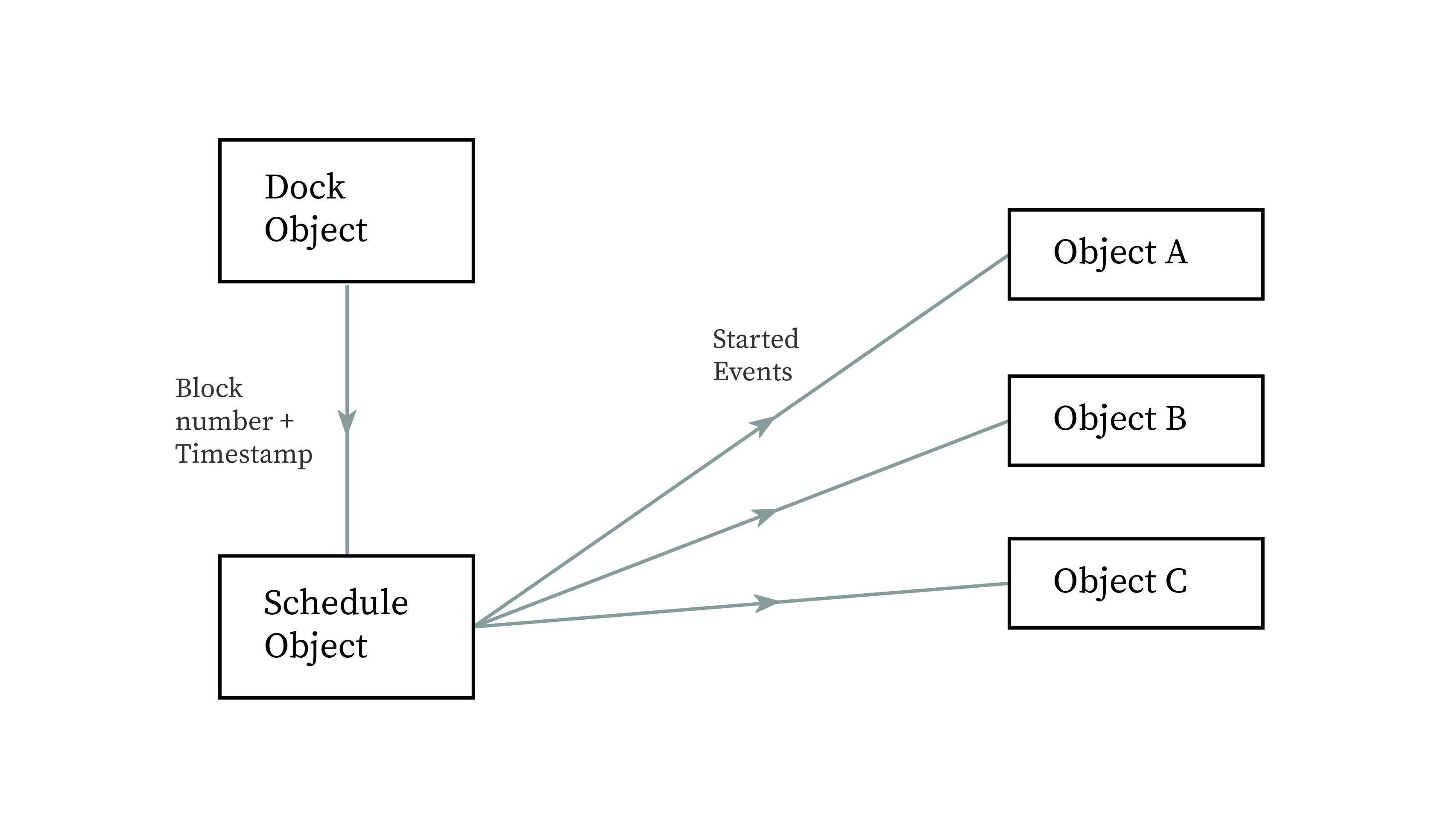}

Every block the dock object sends the current block number and time to the schedule object. The schedule object has a list of functions, IDs and conditions. Upon receiving the block number and time, it goes through this list and, if any of the conditions is satisfied, it triggers the corresponding function at the corresponding ID.

\subsection{Oracles and data feeds}
Each oracle contract only maintains one specific data feed. When they request an update, a Town Crier server can send a transaction containing the update and potentially receive a reward. Other objects can send a function call requesting data from the oracle and it will respond back with the most recent data in its state.

\includegraphics[width=\linewidth]{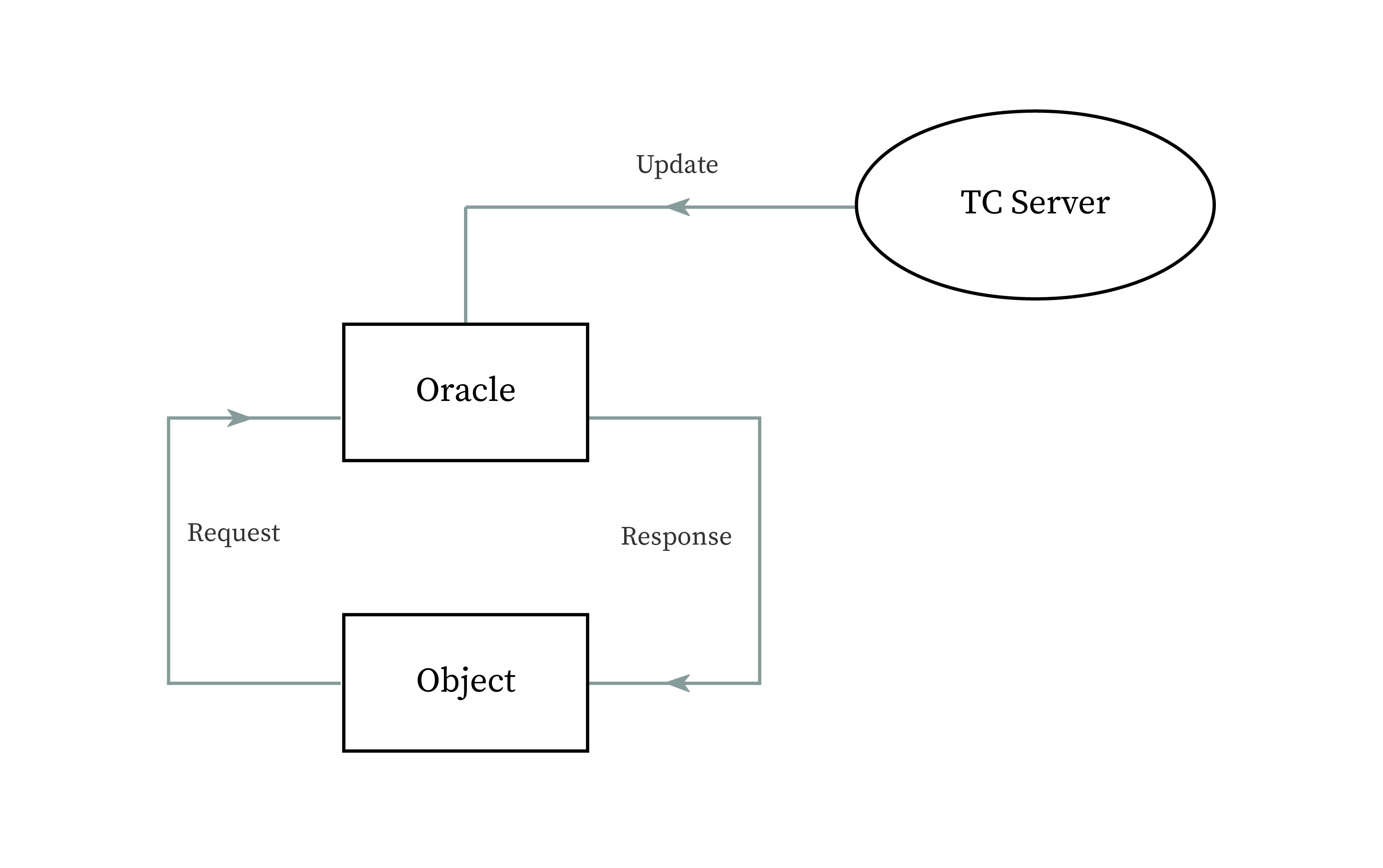}

\subsection{Template instantiation}
Users can create new objects using the instantiation object. Through the dock object, a user can request the instantiation object to create a new object from a list of templates. The instantiation object then sends the necessary events to the super object and also, in some cases, to the authentication and schedule objects.

\includegraphics[width=\linewidth]{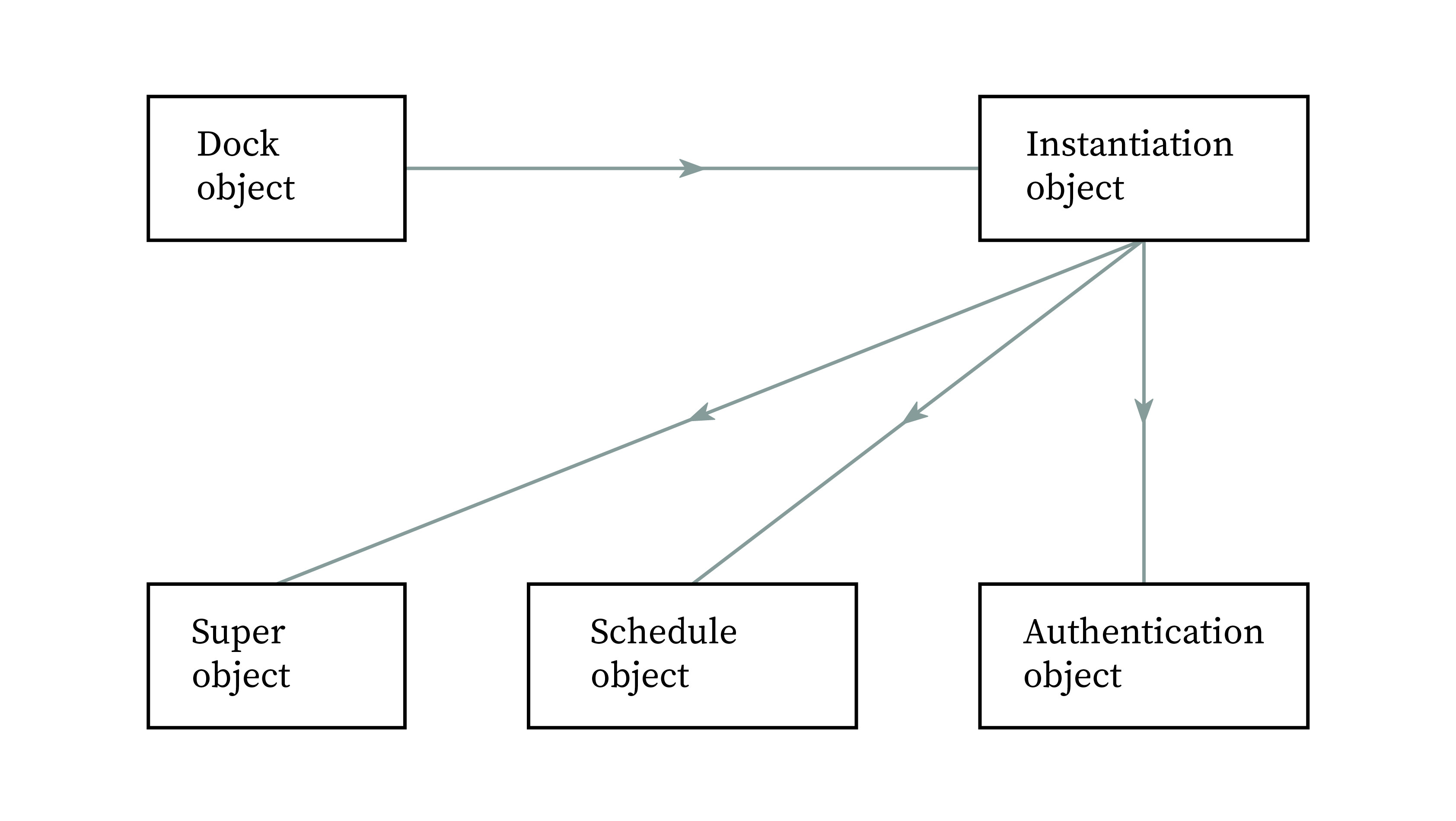}

\subsection{Governance}
The main purpose of the governance object is to control who can have unrestricted access to the state, and especially to the super object. It will receive calls from the dock object and, given its internal logic, it will decide if the calls are to be forwarded or not.

\section{Specification}
In this section we will detail the runtime of Katal, beginning by giving some general remarks and then by describing the kernel and user objects. Our focus will be on the objects interface and function.

\subsection{General}

\subsubsection{Reserved namespace}
There is a reserved \textit{namespace} for the kernel objects. The namespace consists of all the IDs that begin with \textit{'XTL'}. For example, the dock object will have the name \textit{'XTL\_Dock'}.

User objects are not allowed to have an ID that begins with \textit{'XTL'}. This is enforced by the instantiation object, who will not accept any requests to create an user object with an ID in that namespace.

\subsubsection{Transactions}
A transaction is a bundle of function calls to objects. Users can join several different function calls into the same transaction and sign them. The function calls are only forwarded by the dock object if all the signatures are valid. A transaction has the following format:

\begin{itemize}
	\item \textbf{[ID]:} A list of the IDs that are authorizing this transaction.
	\item \textbf{[Function calls]:} A list of the desired function calls.
	\item \textbf{Window:} Determines the time window in which this transaction is valid. For example, if the window is $[2000, 2050]$ then the transaction will only be accepted between the block number 2000 and the block number 2050.
	\item \textbf{Nonce:} A long integer chosen by the users. Together with the validity it prevents \textit{replay attacks}.
	\item \textbf{[Signature]:} A list of the signatures for this transaction, one for each ID.
\end{itemize}

Where [.] represents a list. A transaction can also be thought of as a \textit{wrapper} to a series of function calls. These function calls have the following format:

\begin{itemize}
	\item \texttt{ID\_to:} The ID of the destination object.
	\item \texttt{call\_function:} The function to be called at the destination object.
	\item \texttt{[user\_parameters]:} A list of user-provided parameters to be passed to the function.
\end{itemize}

\subsubsection{Origin ID}
All function calls, except the extrinsics fed into the dock object, will include in their parameters the field \texttt{ID\_from} which is the ID of the object that originated the function call.

Adding this information to every function call allows objects to have functions that can only be called by certain objects. For example:

\begin{itemize}
	\item Internal functions that can only be called by other functions in the same object,
	\item Kernel functions that can only be called by kernel objects,
	\item User-forbidden functions which are functions that do not accept calls from the dock object, thus they cannot be called by users,
	\item User-only functions which are functions that are meant to be called by users and as a result only accept calls from the dock object.
\end{itemize}

\subsubsection{Extrinsics order}
A block is composed of several extrinsics, either transactions or inherents, and they are fed into the dock object into a specific order:

\begin{itemize}
	\item \texttt{Timestamp:=\{time, block number\}:} An inherent that contains timing information. It is forwarded to the schedule object.
	\item \texttt{Slash:=\{ID\}:} An inherent that is used to slash the stake of a misbehaving validator. It is forwarded to the consensus object.
	\item \texttt{Seed:=\{seed\}:} An inherent that contains a random seed used to select a new validator set. It is forwarded to the consensus object only in macro blocks, otherwise it is ignored.
	\item \texttt{Transactions:=\{[transactions]\}:} The list of transactions. They are only sent in micro blocks.
	\item \texttt{Reward:=\{ID\}:} An inherent containing the ID to which the block reward is paid to. It is forwarded to the consensus object.
\end{itemize}

\subsection{Kernel objects}

\subsubsection{Dock}
The dock object acts as the point of entry for extrinsics and its interface only has one function:

\begin{itemize}
	\item \texttt{input(extrinsic):} Parses the extrinsic and verifies its validity. If it is a transaction then it calls the authorization object to verify the signatures and, if the all signatures are valid, then forwards the function calls to the correct objects.
\end{itemize}

For each function call in the extrinsic a new event is created by the dock object with the format \texttt{e(priority, ID\_to, call\_function, ID\_from, [auth\_ID], [user\_parameters])}, where \texttt{[auth\_ID]} is the list of every ID that signed the transaction.

\subsubsection{Authentication}
The authentication object verifies authentication proofs provided by the users. To do so it maintains the following internal key-value store, called the \textit{authentication registry}:

\begin{center}
	\begin{tabular}[c]{c|c}
		Key & Value \\
		\hline
		ID & (Method, [Parameters]) \\
	\end{tabular}
\end{center}

Where method is any authentication method supported by the authentication object. It can be a signature scheme like ECDSA, Schnorr or BLS, or a multisignature scheme, or a hash-lock, or even a zero-knowledge proof system. The parameters are any information necessary to authenticate proofs. For example, in the case of ECDSA the parameters would be the public key.

The next functions form the interface of the authentication object:

\begin{itemize}
	\item \texttt{verify([message, ID, proof]):} Goes through the list verifying that each proof is a valid  authentication of the message by the corresponding ID. Accepts calls from any object.
	\item \texttt{method(message, proof, parameters):} General function type for authentication. There is one instance for each different authentication method. It is an internal function, and as such it accepts only calls from the authentication object.
	\item \texttt{add\_key(ID, method, parameters):} Adds the ID with given method and parameters to the authentication registry. Does not accept calls from the dock object.
	\item \texttt{change\_key([auth\_ID], ID, method, parameters):} Changes the authentication method of the object with given ID. Accepts calls from any object, but if the call originates from the dock object, it will only be accepted if ID $\in$ \texttt{[auth\_ID]}.
	\item \texttt{delete\_key([auth\_ID], ID):} Deletes the entry in the authentication registry with given ID. Accepts calls from any object, but if the call originates from the dock object, it will only be accepted if ID $\in$ \texttt{[auth\_ID]}.
\end{itemize}

\subsubsection{Schedule}
The schedule object serves to make function calls to other objects at regular intervals. To achieve this, it maintains an internal key-value store called the \textit{schedule registry}:

\begin{center}
	\begin{tabular}[c]{c|c}
		Key & Value \\
		\hline
		(ID, Function) & Condition \\
	\end{tabular}
\end{center}

The key identifies which function needs to be called at which object, and the value states under which conditions the function call will be triggered.

Its interface is composed of the following functions:

\begin{itemize}
	\item \texttt{init(block\_number, timestamp):} The main function of the schedule object, it is called once every micro block by an inherent. It goes through the entire schedule registry and checks if block\_number and timestamp satisfies any of the conditions. If it does, it triggers the corresponding function call.
	\item \texttt{add\_key(ID, function, condition):} Adds the given ID, function and condition to the schedule registry. Does not accept calls from the dock object.
	\item \texttt{delete\_key(ID, function):} Deletes the entry in the schedule registry with given ID and function. Does not accept calls from the dock object.
\end{itemize}

\subsubsection{Instantiation}
The instantiation object creates new user objects from a set of object templates. Its interface consists of only two functions:

\begin{itemize}
	\item \texttt{create([template\_ID, parameters]):} For each item in the list it checks if the requested ID is available and, if they are all available, then instantiates the requested objects from \texttt{template\_ID} with the given parameters. Accepts calls from any object.
	\item \texttt{template\_ID(parameters):} General function type, one exists for each object template. It instantiates a new object with the given parameters. It is an internal function.
\end{itemize}

\subsubsection{Governance}
The specific implementation of the governance object will depend on the governance method used. So, it is not possible for us to give a description of the object that will be valid in every case.

However, we will exemplify the simplest case, which is when one single ID has full control over the governance. This hypothetical \textit{dictatorship} object would have the following interface:

\begin{itemize}
	\item \texttt{transmit([auth\_ID], ID\_call, function\_call, parameters):} If \texttt{dictator\_ID} $\in$ \texttt{[auth\_ID]}, it forwards the requested function call. Where \texttt{dictator\_ID} is hard coded into the function. This function basically gives \texttt{dictator\_ID} full access to the entire state.
\end{itemize}

\subsubsection{Native issuance}
The interface for the native issuance object is exactly the same as any other issuance object (see \ref{issuance}), since it is just an instance of the issuance template.

However, we can be more specific regarding its specification:

\begin{itemize}
	\item There is only one \texttt{asset\_ID}, which is the native token \textit{XTL}.
	\item The \texttt{mint} and \texttt{burn} functions only accept calls from the consensus object.
\end{itemize}

\subsubsection{Consensus}
The consensus object deals with all the tasks related to the validators, specifically managing stake deposits, collecting fees and distributing block rewards. To do so, the consensus object has a key-value store of all potential validators, called the \textit{validator registry}:

\begin{center}
	\begin{tabular}[c]{c|c}
		Key & Value \\
		\hline
		ID & (deposit, validating key, status) \\
	\end{tabular}
\end{center}

The key is just the ID of the account that deposited the stake. The value contains the amount deposited (in \textit{XTL}), the public key used to validate blocks and the status. The status just indicates if the validator is currently active or not.

\begin{itemize}
	\item \texttt{stake([auth\_ID], ID, amount, validating\_key):} Transfers the given amount of \textit{XTL} to the consensus object. If successful, it adds ID, validating\_key and amount to the validator registry. It accepts calls from any object, but only if ID $\in$ \texttt{\texttt{[auth\_ID]}}.
	\item \texttt{restake(ID, new\_validating\_key, signature):} Calls the authentication object to check if signature is valid. If it is, it replaces the validating key of ID.
	\item \texttt{unstake([auth\_ID], ID):} Returns deposit back to ID and removes the corresponding key from validator registry. It accepts calls from any object, but only if ID $\in$ \texttt{[auth\_ID]}.
	\item \texttt{fee([auth\_ID], ID, amount):} Transfers a given amount of \textit{XTL} to the consensus object. If successful, updates block reward value. It accepts calls from any object, but only if ID $\in$ \texttt{[auth\_ID]}.
	\item \texttt{slash(ID):} Divides the deposit amount of the given ID between all other active validators and deletes the corresponding key from the validator registry. Can only be called with an inherent.
	\item \texttt{reward(ID):} Transfers the block reward to the given ID. Can only be called with an inherent.
	\item \texttt{change\_validators(seed):} Produces a new list of active validators from the seed and updates the registry accordingly. Can only be called with an inherent.
\end{itemize}

\subsection{User objects}

\subsubsection{Account}
Accounts are the object most utilized by users and also the simplest. They are basically just an ID. All the functionality normally associated with accounts, like transferring tokens and entering into contracts, is provided by other objects.

The interface of an account has a single function:

\begin{itemize}
	\item \texttt{self\_destruct():} Deletes this object and all data associated with it.
\end{itemize}

\subsubsection{Issuance} \label{issuance}
Issuance objects create and destroy tokens, and maintain a list of everyone who owns tokens. Each issuance object can have several different token types, each identified with \texttt{asset\_ID}. Because of this property, issuance objects can support both fungible and non-fungible tokens.

Its interface is composed of the following functions:

\begin{itemize}
	\item \texttt{mint(owner\_ID, asset\_ID, amount):} Creates a given amount of tokens of type \texttt{asset\_ID} and deposits them in \texttt{owner\_ID}. Depending on the option chosen, it can be called by no one, by a predetermined set of IDs or it can be scheduled.
	\item \texttt{burn(owner\_ID, asset\_ID, amount):} Destroys some amount of tokens of type \texttt{asset\_ID} in \texttt{owner\_ID}. Depending on the option chosen, it can be called by no one, by a predetermined set of IDs or it can be scheduled.
	\item \texttt{transfer(owner\_ID, destination\_ID, asset\_ID, amount):} Sends a given amount of tokens of type \texttt{asset\_ID} from \texttt{owner\_ID} to \texttt{destination\_ID}. Accepts calls from any object, but if the call originates from the dock object, it will only be accepted if \texttt{owner\_ID} $\in$ \texttt{[auth\_ID]}.
	\item \texttt{check(owner\_ID, asset\_ID):} Returns the balance in \texttt{asset\_ID} tokens of \texttt{owner\_ID}. Does not accept calls from the dock object.
	\item \texttt{self\_destruct():} Deletes this object and all data associated with it. Depending on the option chosen, it can be called by no one, by a predetermined set of IDs or it can be scheduled.
\end{itemize}

\subsubsection{Oracle}
Oracle objects are the front-end to the Town Crier system. Their main purpose is to maintain a data feed. Such a data feed can be any tuple, as long as it includes a timestamp:

\begin{center}
	\texttt{Data:=\{a, b, ... , timestamp\}}
\end{center}

Oracle objects have the following interface:

\begin{itemize}
	\item \texttt{request():} Turns on a flag stating that it will allow updates to the data feed.
	\item \texttt{update(value, timestamp, proof, receiving\_ID):} Updates the oracle data feed with the given value and timestamp, after verifying the accompanying proof. It can provide a reward to \texttt{receiving\_ID}. It accepts calls from any object, but it may be permissioned.
	\item \texttt{fetch():} Returns the latest value and timestamp.
	\item \texttt{set\_reward(issuance\_ID, asset\_ID, amount):} Sets the reward per update to amount of \texttt{asset\_ID} at the object issuance\_ID. Depending on the option chosen, it can be called by no one, by a predetermined set of IDs or it can be scheduled.
	\item \texttt{self\_destruct():} Deletes this object and all data associated with it. Depending on the option chosen, it can be called by no one, by a predetermined set of IDs or it can be scheduled.
\end{itemize}

\subsubsection{Contract}
Contracts are any financial contracts that are a part of the ACTUS standard. There are 32 different contracts, so it is not possible for us to give a detailed description of every single one. However, the interfaces of all contracts follow the same pattern.

The interface of a contract has two main functions: 1) processing events and 2) managing ownership of the contract. This results in the following general interface:

\begin{itemize}
	\item \texttt{event(parameters):} General function type, it is any event of the associated ACTUS contract type. It can be triggered by a user or by the schedule object. When triggered, the contract will update its internal state, fetch data from an oracle, observe another contract and/or create a cash flow.
	\item \texttt{transfer(owner\_ID, destination\_ID, position, amount):} Transfers fractional ownership of the contract, corresponding to either the creator or counterparty position. Accepts calls from any object, but if the call originates from the dock object, it requires authorization of both \texttt{owner\_ID} and \texttt{destination\_ID}. Depending on the contract parameters, ownership transfer may not be allowed for the creator, the counterparty or both.
	\item \texttt{check(owner\_ID, position):} Returns the ownership amount of a given position by \texttt{owner\_ID}. Does not accept calls from the dock object.
	\item \texttt{self\_destruct():} Deletes this object and all data associated with it. It can be scheduled or triggered by a contract event.
\end{itemize}

Some explanation is needed about the ownership mechanism. In order to support tokenization contracts have some of the functionality of a issuance object. Specifically, all contracts have exactly two different asset IDs, one for the creator position and another for the counterparty position. Also, each of these asset IDs has exactly one token unit. So, if someone holds 0.2 counterparty tokens, that means that he owns 20\% of the counterparty position.

Contrary to the tokens of issuance objects, which are just symbolic representations of external assets, contract tokens have a more active role inside Katal. Contract tokens give whoever holds them a share in the future cash flows (positive or negative) generated by the contract.

A cash flow is executed by a contract by directly calling an issuance object or another contract object. For example, imagine that Alice has 0.2 creator tokens, Bob has 0.8 creator tokens and Charlie has 1 counterparty token. If a contract generates a cash flow of 100 \textit{XTL} from the creator to the counterparty, it will call the \textit{XTL} issuance object to transfer 20 \textit{XTL} from Alice to Charlie and 80 \textit{XTL} from Bob to Charlie.

\section{Functionality}
Developers can build an endless variety of financial services in Katal by combining different ACTUS contracts. In this section we will give a few examples of the financial applications that can be created.

\subsection{Asset-backed tokens}
Issuance objects can be used to create tokens that represent any asset, all that is needed is a trusted entity to hold custody of the underlying assets and to allow the exchange between the token and the asset. This is one of the methods used to create stablecoins.

\includegraphics[width=\linewidth]{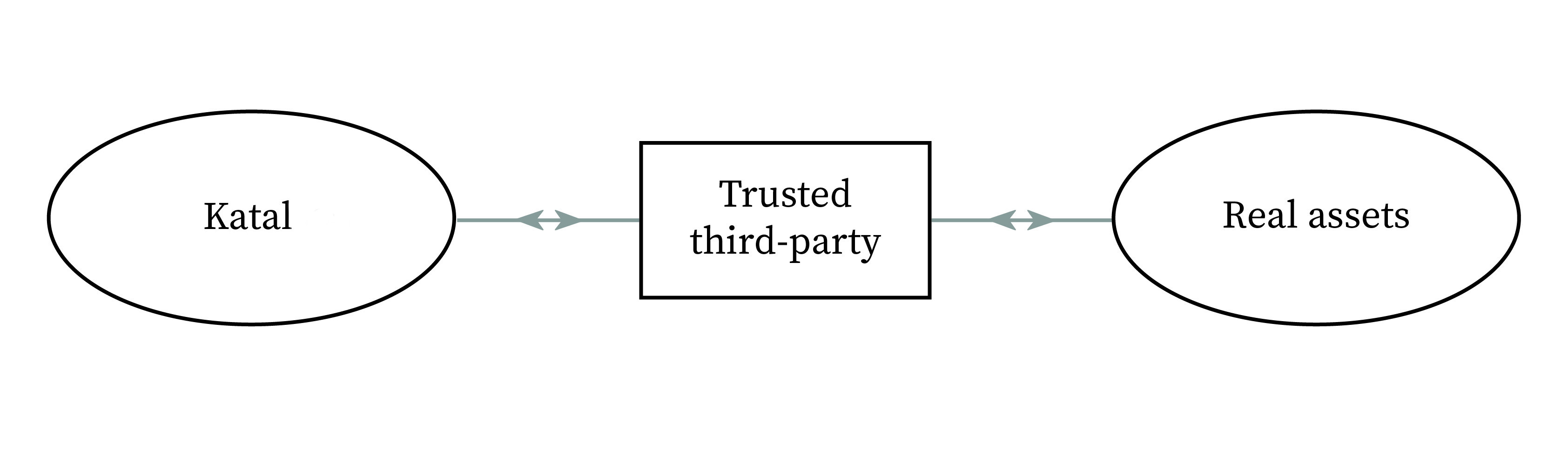}

Any asset can be used to create asset-backed tokens, for example:

\begin{itemize}
	\item Fiat currencies like US Dollar, Euro, Swiss Francs, etc,
	\item Stocks of exchange-traded companies,
	\item Commodities like gold, silver or oil,
	\item Land, houses and other real estate.
\end{itemize}

By far, the most useful asset-backed tokens are the ones backed with fiat, since they can be used as a settlement currency for financial contracts. But other assets also open interesting possibilities, like using stock-backed tokens to create a Katal stock exchange, or using real-estate-backed tokens to create a land registry.

Some critics point out that asset-backed tokens are centralized, but the matter of fact is that there is no better option.

For tangible assets, like commodities and real estate, no blockchain can hold custody of them, so a central entity is required. \textit{Physical assets can not be stored in a blockchain.}

For intangible assets, like stocks and currency, it is technically possible to store them in a blockchain. However, by definition, these are centralized assets. A central bank controls the currency that it issues, and a company controls its own stock.

\subsection{Transfers}
Transfers are the simplest financial service and also one of the most basic functions for any blockchain. The convenience and usability of transfers in Katal matches, or even exceeds, that of banks, online payments systems and other cryptocurrencies:

\begin{itemize}
	\item Variety of currencies: Katal supports transfers of both fiat currencies, through asset-backed tokens, and cryptocurrencies, through Polkadot connectivity.
	\item Human-memorable addresses: Accounts in Katal can have any unique string as their address, thus being more user friendly than other blockchains and banks.
	\item Variety of authentication: Katal supports a variety of authentication methods and lets its users pick the method that they prefer for their own account.
	\item Speed: Katal' consensus algorithm, Albatross, can finalize transactions in just a few seconds.
\end{itemize}

\includegraphics[width=\linewidth]{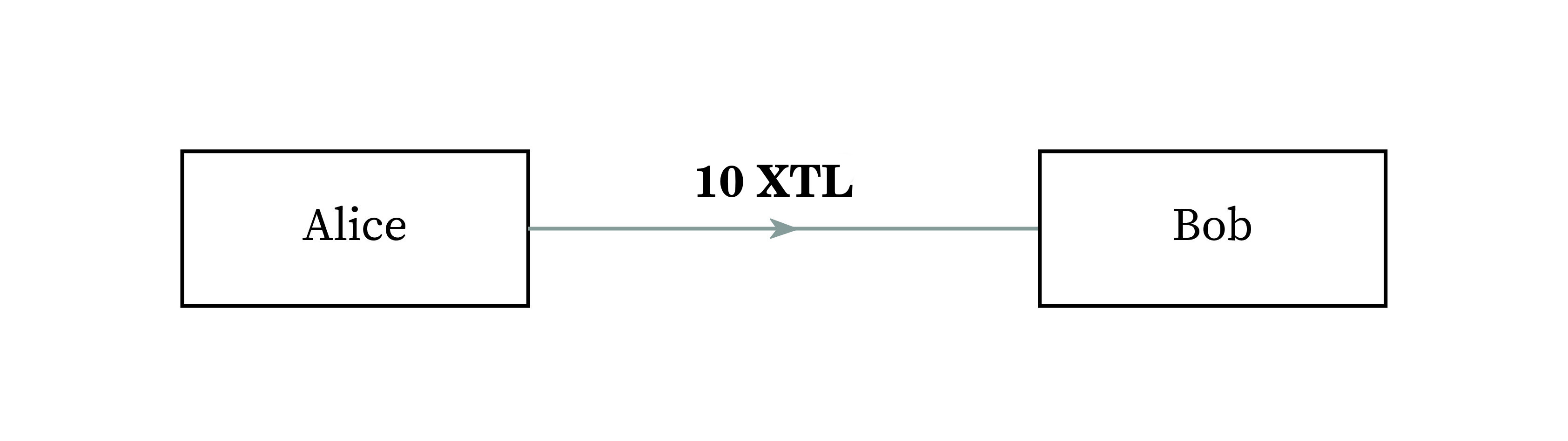}

\subsection{Token exchange}
Many blockchain projects revolve around doing exchanges between different tokens. In Katal, a token exchange can be executed with a single contract, called a \textit{foreign exchange contract}.

\includegraphics[width=\linewidth]{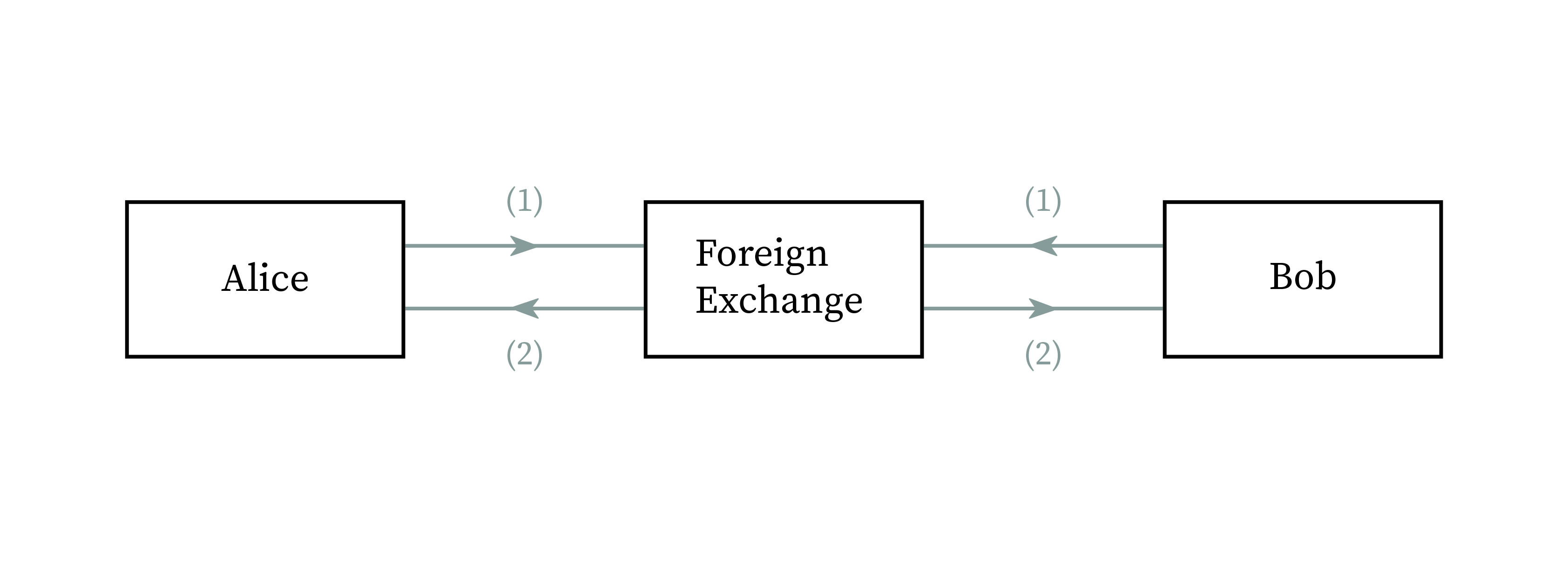}

If Alice and Bob want to exchange two types of tokens, they only need to create the contract. Afterwards, it will transfer the respective tokens out of Alice's and Bob's accounts and transfer them to their new owner.

The foreign exchange contract is \textit{atomic}, meaning that it either completes successfully or it does not happen at all. There is no risk for any of the parties involved.

Given that Katal supports both asset-backed tokens and cryptocurrency tokens (of any blockchain that connects to Polkadot), in addition to its native token \textit{XTL}, the token exchange contract can be used in a variety of interesting situations:

\begin{itemize}
	\item Fiat-XTL: Buying XTL inside Katal using a fiat-backed token.
	\item Fiat-Stock: Buying and selling stocks using a fiat-backed token, akin to a stock exchange.
	\item Fiat-Crypto: Buying and selling cryptocurrencies with fiat.
	\item Crypto-Crypto: Exchanging different cryptocurrencies.
\end{itemize}

\subsection{Collateralized loans}
More complex services can be constructed, for example collateralized loans. Imagine Alice has a house, which is represented in Katal by an asset-backed token, and she wishes to ask Bob for a loan while giving her house as collateral. To do this, Alice and Bob first create two contracts, an annuity contract and a collateral contract, that will codify the terms of the loan.

\includegraphics[width=\linewidth]{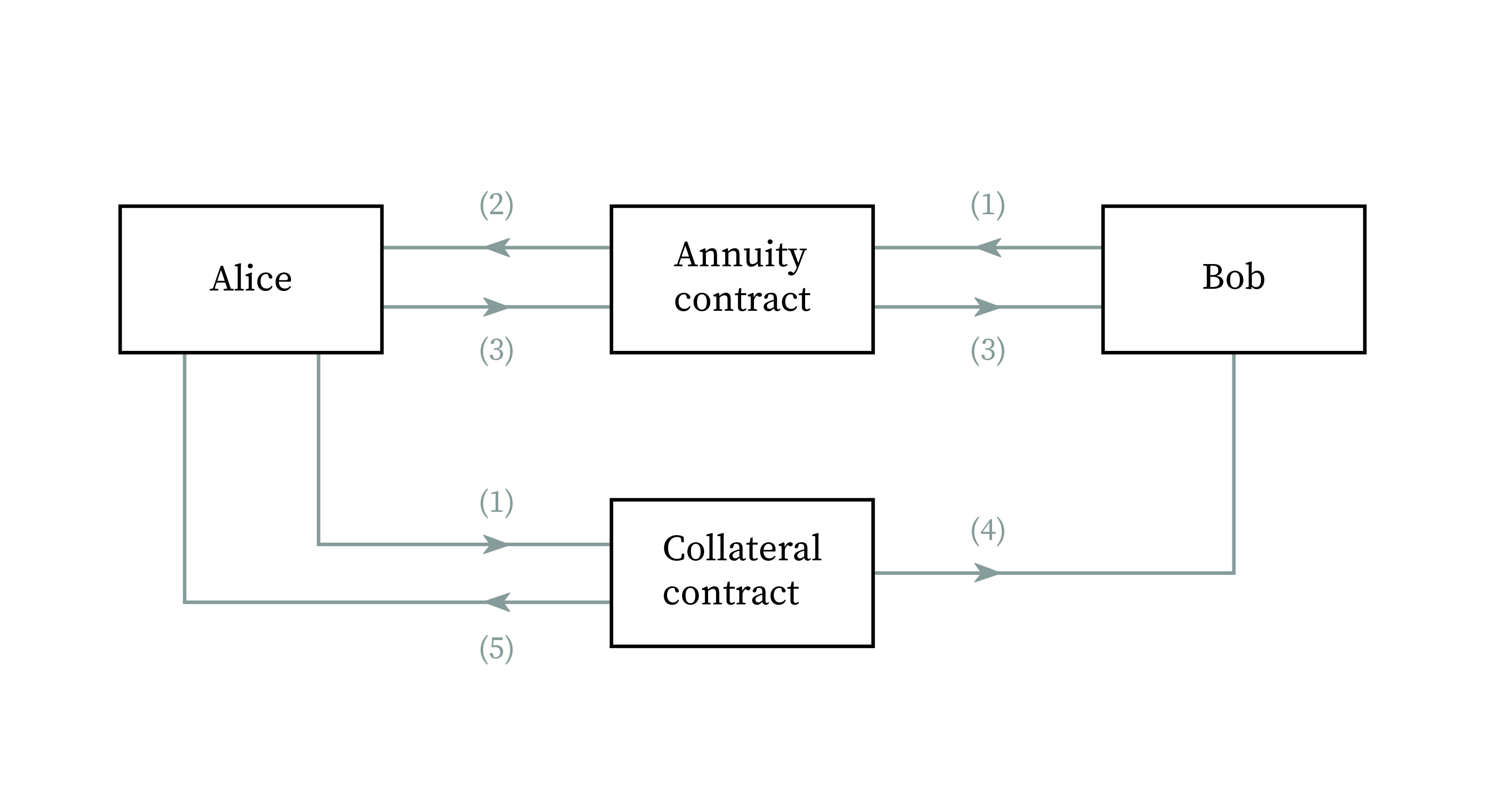}

After creating the contracts, the following series of cash flows will happen:

\begin{enumerate}
	\item The loan amount from Bob's account and the collateral from Alice's account are transferred into the corresponding contracts.
	\item The loan amount is transferred to Alice's account.
	\item Periodically, payments are transferred from Alice's account to Bob's account.
	\item If at any time, Alice does not have enough money in her account for the loan payment, the collateral is transferred to Bob.
	\item If every loan payment is made, at the end of the loan contract the collateral is returned to Alice's account.
\end{enumerate}

Note that Alice and Bob do not need to interact with the blockchain after the contracts are created. All the cash flows are initiated and managed automatically by the contracts.

\subsection{Margin trading}
Margin trading is the act of borrowing money to buy assets and is widely used in all areas of finance.

In margin trading borrowers are required to maintain the net value of their position, meaning the difference between the value of the assets and the value of the loan, at a constant value. This value is called the \textit{margin}.

This financial service can also be done in Katal, and is in fact similar to collateralized loans. Imagine Alice wants to buy \textit{XTL} on margin and to do so she will borrow money, in the form of \textit{XTL} tokens, from Bob. Alice will maintain her margin using fiat-backed tokens. They will create two contracts: a collateral contract and a margin contract.

\includegraphics[width=\linewidth]{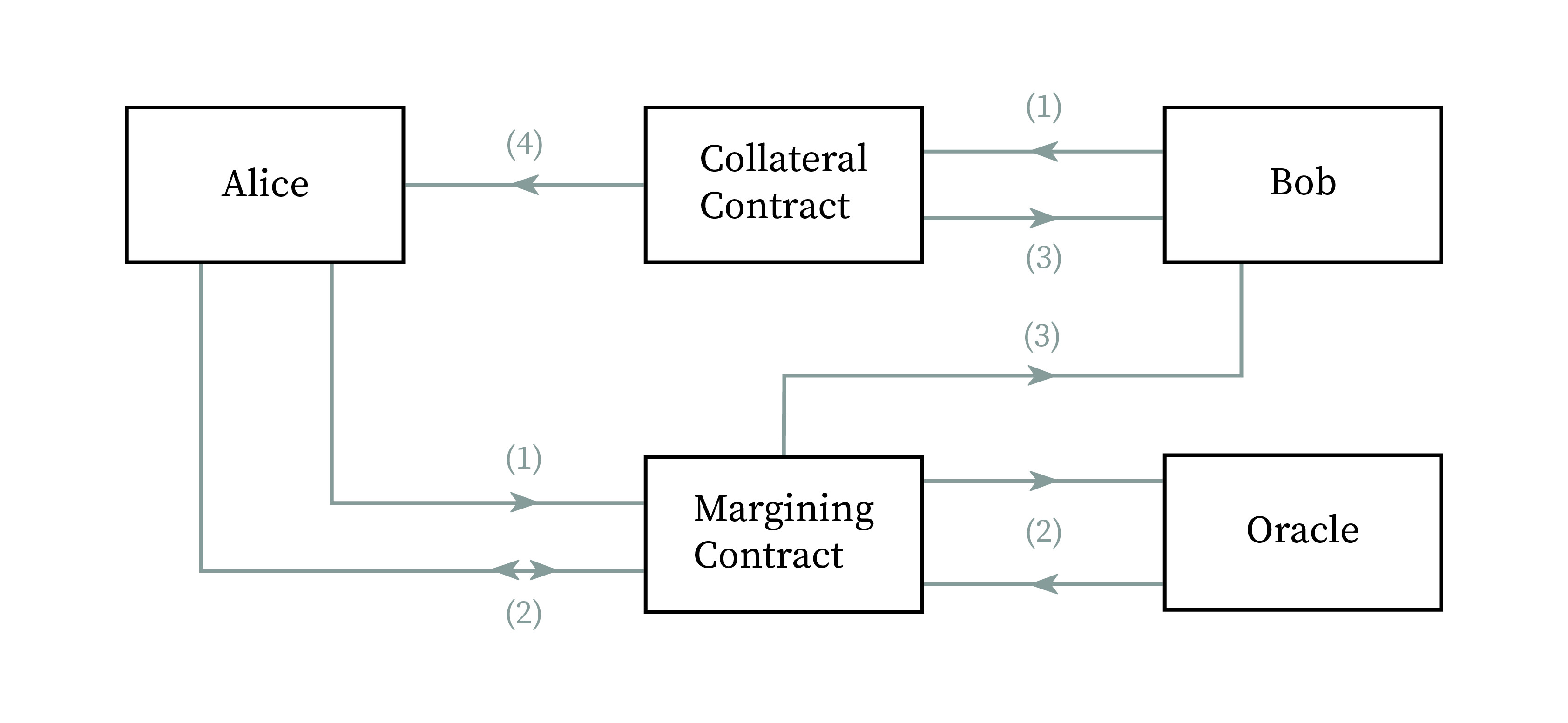}

An oracle for the price of \textit{XTL} is also necessary for adjusting the margin amount. With the contracts created, the following cash flows will take place:

\begin{enumerate}
	\item The borrowed \textit{XTL} is transferred from Bob's account into the collateral contract, and the margin amount is transferred from Alice's account into the margining contract.
	\item Periodically, the margining contract will query the oracle for the \textit{XTL} price and adjust the amount of margin accordingly, transferring cash in and out of Alice's account to maintain the necessary margin.
	\item When Alice decides to terminate her position, or when she is no longer able to maintain the margin, the collateral and the margin are transferred into Bob's account.
	\item If, when her position is terminated, Alice has made a profit then part of the collateral will be transferred into her account.
\end{enumerate}

In this case we exemplified margin buying, but a similar scheme can be used for margin selling, also known as short-selling.

\subsection{Futures}
Futures are one of the most versatile financial contracts, allowing anyone to speculate not just on assets, but on practically anything, for example commodities, stocks, bonds, cryptocurrencies, fiat currencies, indexes, interest rates, energy, weather, etc. As long as there is some publicly available numerical value, a futures contract can be created for it.

Imagine Alice and Bob want to speculate on the price of oil. First, they need to see if an oracle for the price of oil exists in Katal, this is necessary for the contract. Then, Alice and Bob create three contracts: a futures contracts and two margining contracts, one for each of them.

\includegraphics[width=\linewidth]{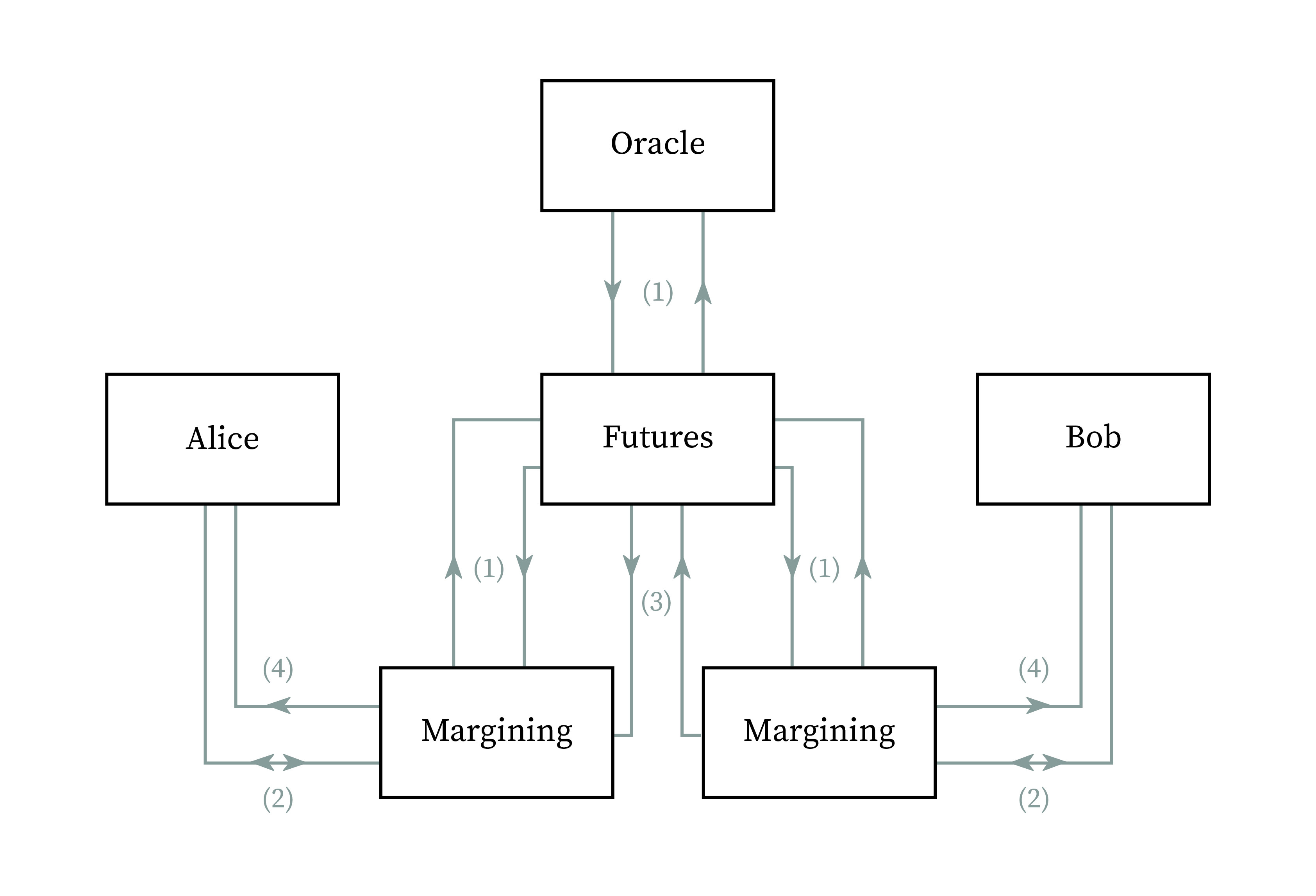}

After the margin amount gets transferred into the margin contracts, the rest of the exchange proceeds as follows:

\begin{enumerate}
	\item At regular intervals, the futures contract fetches the current price of oil from the oracle and updates its state to reflect the profit and loss of each of the parties. The margining contracts also observe the futures contract and adjust the amount of margin required.
	\item The margining contracts transfer cash in and out of Alice's and Bob's account to maintain the necessary amount of margin.
	\item When the futures contract ends, it will settle by making a transfer from one margining contract to the other.
	\item Then the margin contracts will terminate and return their deposits to Alice's and Bob's accounts.
\end{enumerate}

\subsection{Options}
An option is a contract that gives one party the option to buy or sell a given asset at a predetermined price. One party will pay a fee upfront in order to have that option, while the other party will receive that fee. So, the party that buys the option only has upside and the party that sells the option only has downside.

Like futures, options can also be used to speculate on almost any asset. The scheme for options in Katal is also very similar to the futures scheme.

\includegraphics[width=\linewidth]{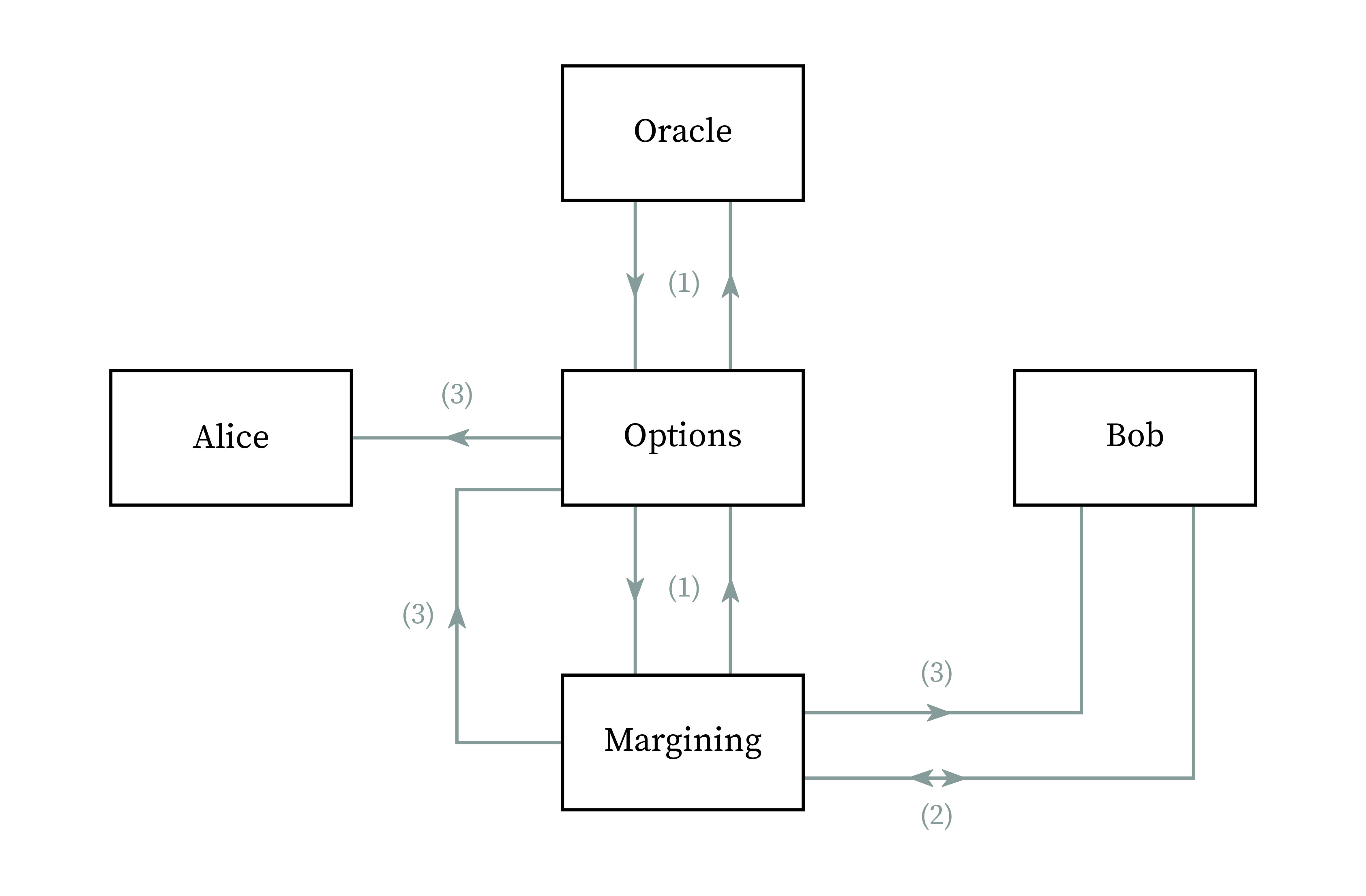}

Imagine Alice wants to buy an option on the price of gold from Bob. Alice will pay Bob for the option and, at the same time, an options contract and a margin contract will be created. The next cash flows will then happen:

\begin{enumerate}
	\item Periodically, the options contract requests the current price of gold from the oracle and updates its state to reflect the profit and loss of each of the parties. The margining contract observes the options contract and adjusts the amount of margin required.
	\item The margining contract transfers cash in and out of Bob's account to maintain the necessary amount of margin.
	\item When Alice exercises the option, or when it expires, it settles by transferring the necessary amount from the margining contract to Alice's account. The rest of the deposit is returned to Bob's account.
\end{enumerate}

\section{Conclusion}
In this paper we introduced Katal, a new blockchain designed purposely for the creation of decentralized financial services and applications. All the different components of Katal allow it to offer a better experience for both users and developers than the one that would be possible using current general-purpose blockchains.

We feel confident that Katal will help revitalize the current decentralized finance industry by making it simpler than ever to create non-custodial trustless interoperable financial contracts.

\section{Acknowledgments}

We would like to acknowledge all other members of the Trinkler Software team, without whom Katal would not be possible. In alphabetical order: Addison Huegel, Arie Levy-Cohen, Herv\'{e} Fulchiron, Mark Greenslade, Nils Bundi and Seraya Takahashi.

\bibliographystyle{IEEEtran}
\bibliography{Katal.bib}

\end{document}